\documentclass[aps,prc,preprint,groupedaddress,showpacs,preprintnumbers,amsmath,amssymb]{revtex4}

\usepackage{graphicx}
\usepackage{dcolumn}
\usepackage{bm}

\begin{document}

\preprint{PRC}

\title{Two-body Photodisintegration of $^{4}$He \\with Full Final State Interaction}


\author{Sofia Quaglioni, Winfried Leidemann and Giuseppina Orlandini}
\affiliation{Dipartimento di Fisica, Universit\`a di Trento, and Istituto
Nazionale di Fisica Nucleare, Gruppo Collegato di Trento, I-38050 Povo, 
Italy}

\author{Nir Barnea}
\affiliation{The Racah Institute of Physics, The Hebrew University, 
91904, Jerusalem, Israel}

\author{Victor D. Efros}
\affiliation{Russian Research Centre ``Kurchatov Institute'', Kurchatov 
Square, 1, 123182 Moscow, Russia}


\date{\today}

\begin{abstract}
The cross sections of the processes $^4$He($\gamma,p$)$^3$H and 
$^4$He($\gamma,n$)$^3$He are calculated taking into account the full final 
state interaction via the Lorentz integral transform (LIT) method.  This 
is the first consistent microscopic calculation beyond the three--body 
breakup threshold. The results are obtained with a semirealistic central 
NN potential including also the Coulomb force. The cross sections show a 
pronounced dipole peak at 27 MeV which lies  within the rather  broad  
experimental band. 
At higher energies, where experimental uncertainties are considerably
smaller, one finds a good agreement  between theory and experiment. 
The calculated sum of three-- and  four--body photodisintegration cross 
sections is also listed and is in fair agreement with the data.  
\end{abstract}


\maketitle


\section{Introduction}
The  photodisintegration of the $^{4}$He nucleus has a longstanding 
history. 
First experiments on the $(\gamma,n)$ reaction were performed some 50 
years ago \cite{Fer54}. In the following 25 years various experimental 
groups carried out measurements for both the $^4$He$(\gamma,p)^3$H and 
$^4$He$(\gamma,n)^3$He reaction channels and the inverse capture reactions 
\cite{Cle65,Den68,Gor68,Mey70,Dod72,Iri73,Mal73,Ark74,Iri75,Bal77,Ber80,War81,McB82,Cal83}. 
Dramatically conflicting results have 
been obtained as a result of this work. The $(\gamma,p)$ data were  
consistent in showing a rather pronounced resonant peak close to the 
three--body breakup threshold. 
At the same time, the $(\gamma,n)$ data at low energy were very spread, 
and measurements showed  either a strongly pronounced  or a rather 
suppressed giant dipole peak. In 1983 a careful and balanced review of 
all the available experimental data for the two mirror reactions was 
provided \cite{CBD83}.
A strongly peaked cross section at low energy  for the ($\gamma,p$) 
channel and a flatter shape for the ($\gamma,n$) one was recommended 
by the authors. Three new experiments on the ($\gamma,p$) reaction 
were subsequently carried out, two of them contradicting and one 
confirming the recommended cross section. In particular in \cite{Ber88} 
and \cite{Fel90} a suppressed dipole resonance was found, whereas 
\cite{Hoo93} proved to be   
in agreement with the previous strongly peaked results. Measurements of 
the ratio of the $(\gamma,p)$ to  $(\gamma,n)$  cross sections in the 
giant resonance region were  performed  as well and, at variance with 
the cross sections recommended in \cite{CBD83}, results very close to 
unity were reported \cite{Phi79,Flo94}. Finally, in 1992 additional 
cross section data were deduced from a Compton scattering experiment on 
$^{4}$He \cite{Wells92}. A strongly peaked cross section for the 
$^{4}$He total photoabsorption was found suggesting a ($\gamma,n$) 
cross section considerably larger  than the one recommended in 
\cite{CBD83}. 
 
In  1996 the first theoretical calculation  of the two--fragment breakup 
cross section with inclusion of final state interaction 
(FSI) was performed in the energy range below the three--fragment 
breakup threshold \cite{Ell96}. The semi--realistic MTI-III potential 
\cite{MT} was employed. The results of \cite{Ell96} showed a rather 
suppressed giant dipole peak. The following  year a calculation of the 
total $^{4}$He photoabsorption cross section up to the pion threshold
was carried out \cite{ELO97}. Full FSI was taken into account in 
the whole energy range via the Lorentz 
integral transform (LIT) method \cite{ELO94} that does not require 
calculating continuum wave functions. The four--nucleon dynamics was 
described with the same NN potential as in \cite{Ell96}. Different from 
the previous work  a pronounced giant dipole peak was found \cite{ELO97}. 
These results have been reexamined in 
\cite{BELO01}. A small shift of the peak position has been obtained, 
but the pronounced peak has been confirmed. 

In order to understand the origin of the large total photonuclear 
cross section in terms of the various channel contributions, 
a separate calculation of the channel cross sections is necessary.
The present paper reports a theoretical study of the two--fragment 
breakup processes $^{4}$He$\left(\gamma,p\right)^{3}$H and 
$^{4}$He$\left(\gamma,n\right)^{3}$He. For these exclusive reactions 
we investigate the issue  of the giant dipole peak height, and we also 
calculate the cross section at higher energies, up to the pion threshold. 
The calculation includes full FSI via the LIT method \cite{Ef85} and 
employs the semi-realistic MTI-III potential \cite{MT}. 

The LIT method has already been successfully tested for exclusive 
reactions in the $d(e,e^{\prime}p)n$ process \cite{LL00}. The present
calculation represents the first application to $A>2$.
We would like to emphasize 
that in our study FSI is taken into account rigorously also beyond the 
three--and four--body breakup thresholds. In addition, combining our cross 
sections with the total photoabsorption cross section calculated in 
\cite{ELO97,BELO01} for the same potential, we obtain the sum of 
three-- and four--body photodisintegration cross sections of $^4$He, 
for which some experimental data are available 
\cite{Gor58,Ark70,Bal79,Dor93}.   

\section{general formalism}

\subsection{Cross Section}

The total exclusive cross section of the $^4$He photodisintegration 
into the two fragments {\it N} and {\it 3}, where {\it N} refers to 
the scattered proton (neutron) and {\it 3} refers to the $^3$H ($^3$He) 
nucleus, is given by 
\begin{equation}
\sigma_{N,3}=
\frac{e^2}{\hbar c}\frac{k\mu~\! \omega_{\gamma}}{4\pi}
\int{\left|\left\langle\Psi_{N,3}^-(E_{N,3})
\left|D_z\right|\Psi_{\alpha}\right\rangle\right|^2 
d\Omega_k~,~~~ E_{N,3}=\omega_{\gamma}+E_{\alpha}}~.
\label{sigma}
\end{equation}
In the equation above we neglect the very small nuclear recoil energy. 
With $\mu$ and $k$ we denote the reduced mass and the relative momentum 
of the  fragments, respectively, $\omega_{\gamma}$ is the incident 
photon energy, $\Psi_{\alpha}$ is the ground--state wave function, 
$\Psi^-_{N,3}$ is the final--state continuum wave function 
of the minus type  pertaining to the {\it N},{\it 3} channel 
\cite{Gol64}, and $E_{\alpha}$ and $E_{N,3}$ are the energies 
of the corresponding initial and final states, respectively. As in 
\cite{ELO97,BELO01} only  transitions induced by the unretarded dipole 
operator,
\begin{equation}
D_z =\sum_{j=1}^{4}\tau_{j}^{3}z_{j} \,
\label{dipole}
\end{equation}
are taken into account. Here $\tau^3_j$ denotes the third component 
of the $j$-th nucleon isospin and $z_j$ represents the $z$ component of 
the distance between the $j$-th nucleon and the center of mass of the 
system. This form of the transition  operator already includes the 
leading effects of meson exchange currents via Siegert's theorem. 
Additional contributions to the total cross section are small at the
energies considered here (see also \cite{AS91,Gol02}).

The main difficulty in the calculation of such a cross section is the 
presence of the  continuum  wave function $\Psi^-_{N,3}
(E_{N,3})$ in the transition matrix element 
\begin{equation}
T_{N,3}(E_{N,3})=
\left\langle\Psi_{N,3}^-(E_{N,3})\left|D_z\right|
\Psi_{\alpha}\right\rangle~.
\label{tn3}
\end{equation}
As it will be explained in the next subsection, with the LIT method 
one is able to perform an {\it ab initio} calculation of this   
transition matrix element in the whole energy range below the pion 
threshold without dealing with the continuum solutions of the four--body 
Schr\"odinger equation.

\subsection{The LIT Method for Exclusive Reactions}

For the case of an exclusive perturbation--induced process, all the 
information about the reaction dynamics is contained in the transition 
matrix element of the perturbation $\widehat{O}$ between the initial 
($\Psi_0$) and final ($\Psi^-_f$) states, 
\begin{equation}
T_{f}(E_f)=\left\langle\Psi^-_f(E_f)\left|{\widehat O}\right|
\Psi_0\right\rangle~.
\label{tfi}
\end{equation} 
The calculation of such a matrix element can be carried out with the 
LIT method as outlined in the following \cite{Ef85,LL00}.
For simplicity we restrict our discussion to an exclusive reaction 
leading to a final state with two fragments, a fragment $a$ with $n_a$ 
nucleons and a fragment $b$ with $n_b=A-n_a$ nucleons, though more complex 
fragmentations of the initial $A$--body system can be treated as well. 
Denoting  with $H$ the full nuclear Hamiltonian we have a formal 
expression for $\Psi^-_{f=a,b}$ in terms of the "channel state" 
$\phi_{f=a,b}^-(E_{f=a,b})$ \cite{Gol64},
\begin{equation}
\left|\Psi^-_{a,b}(E_{a,b})\right\rangle={\widehat{\mathcal A}}
\left|\phi_{a,b}^-(E_{a,b})\right\rangle+\frac{1}{E_{a,b}-i\varepsilon-H}
\widehat{\mathcal A}{\mathcal V}\left|\phi_{a,b}^-(E_{a,b})\right\rangle,
\label{psi-}
\end{equation}
where $\widehat{\mathcal A}$ is an antisymmetrization operator.
When there is no Coulomb interaction in the system, the channel state
 $\phi_{a,b}^-(E_{a,b})$ is the product of the internal wave 
functions of the fragments 
and of their relative free motion. Correspondingly, ${\mathcal V}$ in 
Eq. (\ref{psi-}) is the sum of all interactions between particles 
belonging to different fragments.  In the general case, 
$\phi_{a,b}^-(E_{a,b})$ is chosen to account for the average Coulomb 
interaction between the fragments, and the plane wave describing their
relative motion is replaced by the Coulomb function of the 
minus type. We write down this function in the partial wave  
expansion form 
\begin{equation}
\phi^-_{a,b}(E_{a,b})=\frac{\Phi_a(1, ...,n_a)\Phi_b(n_a+1, ...,A)}
{(2\pi)^{3/2}}~4\pi\sum_{\ell=0}^{\infty}
\sum_{m=-\ell}^{\ell}{i^{\ell}e^{-i\delta_{\ell}(k)}\frac{w_{\ell}
(k;r)}{kr}Y_{\ell m}(\Omega_r)Y^*_{\ell m}(\Omega_k).}
\label{phiump}
\end{equation}
Here $\Phi_a(1, ..., n_a)$ and $\Phi_b(n_a+1, ..., A)$ are the internal 
wave functions of the the fragments, 
${\mathbf r}=(r,\Omega_r)={\mathbf R}^a_{cm}
-{\mathbf R}^b_{cm}$ represents the distance between them, and the 
energy of the relative motion is $k^2/ 2\mu=E_{a,b}-E_a-E_b$, where $E_a$ 
and $E_b$ are the fragment ground state energies. The functions $w_{\ell}(k;r)$ 
are the regular Coulomb wave functions of order $\ell $, and 
$\delta_{\ell}(k)$ are the Coulomb phase shifts \cite{Gol64}.   

The internal wave functions of the fragments are assumed to be 
antisymmetrized and normalized to unity, so that the properly normalized 
continuum wave function in Eq. (\ref{psi-}) is obtained via application 
of the antisymmetrization operator
\begin{equation}
{\widehat{\mathcal A}}=\sqrt{\frac{n_a!n_b!}{A!}}\left[1-\sum_
{i=1}^{n_a}\sum_{j=n_a+i}^{A}{\mathcal P}_{ij}\right]~,
\end{equation}    
where ${\mathcal P}_{ij}$ are particle permutation operators \cite{Gol64}. 

When one inserts Eq. (\ref{psi-}) into Eq. (\ref{tfi})  the transition 
matrix element becomes the sum of two pieces, a Born term,
\begin{equation}
T_{a,b}^{Born}(E_{a,b})=\left\langle\phi_{a,b}^-(E_{a,b})\left|
{\widehat{\mathcal A}}~{\widehat O}\right|\Psi_0\right\rangle~,
\end{equation}
and an FSI dependent term, 
\begin{equation}
T_{a,b}^{FSI}(E_{a,b})=\left\langle\phi_{a,b}^-(E_{a,b})\left|
{\mathcal V}{\widehat{\mathcal A}}\frac{1}{E_{a,b}+i\varepsilon-H} 
{\widehat O}\right|\Psi_0\right\rangle~.
\label{FSI}
\end{equation}
While the Born term is rather simple to deal with, the determination  
of the FSI dependent matrix element is rather complicated. We treat this 
term within the LIT approach  as outlined in the following. 

Let $\Psi_\nu(E_\nu)$
be the eigenstates of the Hamiltonian labelled  by channel quantum 
numbers $\nu$ and normalized as $\langle\Psi_{\nu}|\Psi_{\nu^{\,\prime}}\rangle=\delta(\nu-\nu^{\,\prime})$. Using the completeness relation of the set 
$\Psi_\nu(E_\nu)$ the matrix element $T_{a,b}^{FSI}(E_{a,b})$ can be 
written as: 
\begin{eqnarray}
T_{a,b}^{FSI}(E_{a,b})&=&\sum\!\!\!\!~\!\!\!\!\!\!\int d\nu\,
\langle\phi_{a,b}^-(E_{a,b})|{\mathcal V}
{\widehat{\mathcal A}}|\Psi_\nu(E_\nu)\rangle\frac{1}{E_{a,b}+i
\varepsilon-E_\nu}\langle\Psi_\nu(E_\nu)| 
{\widehat O}|\Psi_0\rangle \nonumber\\
&=&\int_{E_{th}^-}^{\infty}{\frac{F_{a,b}(E)}{E_{a,b}+i\varepsilon-E} 
dE}~=~-i\pi F_{a,b}(E_{a,b})+{\mathcal P}\int_{E_{th}^-}^{\infty}
{\frac{F_{a,b}(E)}{E_{a,b}-E}dE}~,
\end{eqnarray}
where $F_{a,b}(E)$ is defined as
\begin{equation}
F_{a,b}(E)=\sum\!\!\!\!~\!\!\!\!\!\!\int d\nu\left\langle\phi_{a,b}^-
(E_{a,b})\left|{\mathcal V}{\widehat{\mathcal A}}\right|\Psi_\nu(E_\nu)
\right\rangle\left\langle\Psi_\nu(E_\nu)\left|{\widehat O}\right|\Psi_0
\right\rangle\delta(E-E_\nu)~,
\end{equation}
and $E_{th}$ is the lowest excitation energy in the system i.e. the 
breakup threshold energy. To calculate $T_{a,b}$ one needs to know the 
function $F_{a,b}$ for all energy values. The direct calculation of 
$F_{a,b}$ is of course far too difficult, since one should know all the 
eigenstates $\Psi_\nu$ for the whole eigenvalue spectrum of $H$. However, 
an indirect calculation of $F_{a,b}$ is possible applying the LIT method. 
To this end, one introduces  an integral transform of $F_{a,b}$ with 
a kernel of Lorentzian shape,
\begin{eqnarray}
L\left[F_{a,b}\right](\sigma)&=&\int_{E_{th}^-}^{\infty}
{\frac{F_{a,b}(E)}
{(E-\sigma_R)^2+\sigma_I^2}~dE}~=~\int_{E_{th}^-}^{\infty}{\frac{F_{a,b}
(E)}{(E-\sigma)(E-\sigma^*)}~dE}\nonumber\\
&=&\sum\!\!\!\!~\!\!\!\!\!\!\int d\nu{\left\langle\phi_{a,b}^-(E_{a,b})
\left|{\mathcal V}
{\widehat{\mathcal A}}\frac{1}{H-\sigma^*}
\right|\Psi_\nu(E_\nu)\right\rangle\left\langle\Psi_\nu(E_\nu)\left|
\frac{1}{H-\sigma}{\widehat O}\right|\Psi_0\right\rangle}\nonumber\\
&=&\left\langle {\widetilde \Psi}_2(\sigma)\left|\right.{\widetilde
\Psi}_1(\sigma)\right\rangle~,
\label{psi21}
\end{eqnarray}
with $\sigma=\sigma_R+i\sigma_I$. Here we denote 
\begin{equation} 
{\widetilde\Psi}_1(\sigma)=(H-\sigma)^{-1}{\widehat O}\left|\Psi_0
\right\rangle,\qquad
{\widetilde\Psi}_2(\sigma)=(H-\sigma)^{-1}{\widehat {\mathcal A}}
{\mathcal V}
|\phi_{a,b}^-(E_{a,b})\rangle~.\label{psitdef}
\end{equation}
Equation (\ref{psi21}) shows that $L\left[F_{a,b}\right](\sigma)$ can 
be calculated  without explicit knowledge of $F_{a,b}$ provided that 
one solves the two equations, 
\begin{eqnarray}
(H-\sigma)\left|{\widetilde\Psi}_1\right\rangle&=&{\widehat O}
\left|\Psi_0\right\rangle\label{psitilde1}~,\\
(H-\sigma)\left|{\widetilde\Psi}_2\right\rangle&=&{\widehat 
{\mathcal A}}{\mathcal V}|\phi_{a,b}^-(E_{a,b})\rangle
\label{psitilde2}
\end{eqnarray}
which differ in the source terms only. 

It can be seen that ${\widetilde\Psi}_1$ and ${\widetilde\Psi}_2$ are 
localized i.e. have finite norms. Assuming that $\sigma_I\ne 0$ this 
is ensured by the fact that the source terms in the right--hand sides 
of Eqs. (\ref{psitilde1}) and (\ref{psitilde2}) are localized. In fact
in the coordinate representation the source term in Eq. (\ref{psitilde1}) 
decreases exponentially for increasing distances between
particles. The source term in Eq. (\ref{psitilde2}) contains two potential
contributions i.e. the nuclear force component and the Coulomb interaction.
The contribution due to the nuclear interaction 
vanishes exponentially for increasing distances between particles.
Since the average Coulomb potential between the fragments is already 
taken into account via Eq. (\ref{phiump}), the remaining Coulomb 
contribution behaves like $r^{-2}$, as the distance $r$ from the fragments 
increases. 
Therefore the source term remains localized also in the presence of the 
Coulomb force. In our calculation this Coulomb term arises in case 
of the p,$^3$H channel and its contribution to the results is found 
to be very small. 

In case of breakups into more than two fragments it is convenient 
to write Eq. (\ref{psi21}) in another form:
\begin{eqnarray} 
L\left[F_{a,b}\right](\sigma)&=&\frac{1}{2i\sigma_I}\left\langle
\phi^-(E_{a,b})\left|{\mathcal V}{\widehat{\mathcal A}}\left[\frac{1}
{H-\sigma}-\frac{1}{H-\sigma^*}\right]{\widehat O}\right|\Psi_0
\right\rangle
\nonumber\\
&=&\frac{1}{2i\sigma_I}\left[\left\langle\phi_{a,b}^-(E_{a,b})\right|
{\mathcal V}{\widehat{\mathcal A}}\left|{\widetilde\Psi}_1(\sigma)
\right\rangle-\left\langle\phi_{a,b}^-(E_{a,b})\right|{\mathcal V}
{\widehat{\mathcal A}}\left|{\widetilde\Psi}_1^{\prime}(\sigma)
\right\rangle\right]~.
\label{mod}
\end{eqnarray}
Here ${\widetilde\Psi}_1(\sigma)$ is the same as above,
and ${\widetilde\Psi}_1^{\prime}(\sigma)$ satisfies
\begin{equation}
(H-\sigma^*)\left|{\widetilde\Psi}_1^{\prime}(\sigma)\right\rangle
={\widehat O}\left|\Psi_0\right\rangle~.
\label{psitilde3}
\end{equation}
Therefore, also in this case, one has to deal with a localized solution.

When solving Eqs. (\ref{psitilde1}), (\ref{psitilde2}) and (\ref{psitilde3}) 
it is sufficient to require that the solutions are localized, and no 
other boundary conditions are to be imposed. Therefore, similar to what 
is done in bound state calculations, one can use  an expansion over 
a basis set of localized functions. A convenient choice is the basis set 
consisting of correlated hyperspherical harmonics
(CHH) multiplied by hyperradial functions. 
   
As discussed in \cite{BELO01} for the case of the total $^4$He 
photoabsorption cross section, special attention has to be paid to the 
convergence of such expansions. A rather large number of basis states 
is necessary in order to reach convergence, thus leading to large 
Hamiltonian matrices. Instead of using a time consuming inversion method 
we directly evaluate the scalar products in (\ref{mod}) with the Lanczos 
technique \cite{Mar02}. Inserting  the Lanczos orthonormal basis 
$\{\varphi_i,i=0, ..., n\}$ into Eq. (\ref{mod}), where $\varphi_o$ is 
taken to be the right--hand side (normalized to unity)
of Eqs. (\ref{psitilde1}), (\ref{psitilde3}), one gets 
\begin{equation}
L[F_{a,b}](\sigma)=\frac{\sqrt{\left\langle\Psi_0\left|{\widehat 
O^{\dagger}}
{\widehat O}\right|\Psi_0\right\rangle}}{2i\sigma_{I}}\sum_{i=0}^n
{\left\langle\phi_{a,b}^-(E_{a,b})\left|{\mathcal V}{\widehat{\mathcal A}}\right|
\varphi_i\right\rangle\left\langle\varphi_i\left|\frac{1}{H-\sigma}-
\frac{1}{H-\sigma^*}\right|\varphi_o\right\rangle}~.
\label{lit}
\end{equation}
The matrix elements $\langle\varphi_i|(H-\sigma)^{-1}|\varphi_o\rangle$ 
can be written as continued fractions of the Lanczos coefficients.   

After having calculated $L[F_{a,b}](\sigma)$ one obtains the function 
$F_{a,b}(E)$, and thus $T_{a,b}(E_{a,b})$, via the inversion of the LIT, 
as described in \cite{ELO99}.

\section{results} 

The ground states of $^4$He, $^3$He and $^3$H as well as the LIT in Eq.
(\ref{lit}) are calculated using the CHH expansion method. In order to
speed up the convergence, state independent correlations are introduced 
as in \cite{ELO297}. We use the MTI-III
\cite{MT} potential and identical CHH expansions for the ground state 
wave functions of $^4$He and of the three--nucleon systems as in 
\cite{BELO01} and \cite{ELO97B}, respectively.

Because of the choice of the excitation operator (see Eq. (\ref{dipole})), 
in solving Eqs. (\ref{psitilde1}) and (\ref{psitilde2}) the HH
must be  characterized by the quantum 
numbers L=1, S=0 and T=1. In this calculation the maximal value of the 
grand-angular quantum number, $K_{max}$, is equal to 29. 
Such a high value of 
$K_{max}$ has been made possible by neglecting states which have proved 
to give very small contributions to the LIT. The choice of these states
has been done in the following way. 
Our hyperspherical harmonics (HH) depend on three Jacobi vectors 
${\boldsymbol\xi}_1,{\boldsymbol{\xi}}_2,{\boldsymbol{\xi}}_3$, 
among which $\xi_1$ represents the distance between a pair of particles.
The HH entering the calculation are obtained via symmetrization 
of the HH possessing a definite relative orbital 
momentum $\ell_1$ associated with ${\boldsymbol\xi}_1$ and a definite 
grand--angular momentum $\widetilde{K}$ in the subspace of vectors 
${\boldsymbol\xi}_2,{\boldsymbol\xi}_3$. We have found that 
beyond $K_{max}=13$ we can neglect all symmetrized HH which originate 
from the non-symmetrized 
HH with $\ell_1$ greater than 2 and 
$\widetilde{K}$ greater than 3 and, beyond $K_{max}=19$, 
those with $\ell_1$ 
greater than 0 and $\widetilde{K}$ greater than 1.
This selection is similar to that justified and used in 
non--correlated HH bound 
state calculations (see \cite{FE81} and references therein). 
An additional selection has been 
performed with respect to the permutational symmetry types of the 
HH. For $S=0$, $T=1$ quantum numbers, 
HH of two permutational symmetry types enter 
the expansion, those belonging to the irreducible representations 
$[f]=[+]$ and $[f]=[-]$ of the four--particle permutation group 
${\mathbf S_4}$ \cite{ELO297,FE81}. For basis states constructed 
from the HH of the $[+]$ type the probability 
for a nucleon pair to be in a relative even state and in particular 
in the $s$--state is substantially higher. We have found that
the contribution to the LIT of states with the spatial symmetry $[-]$ 
is suppressed, and the corresponding HH with 
$K$ values higher than 9 can be neglected in the calculation.  
       
\begin{figure}
\resizebox*{14cm}{9cm}{\includegraphics{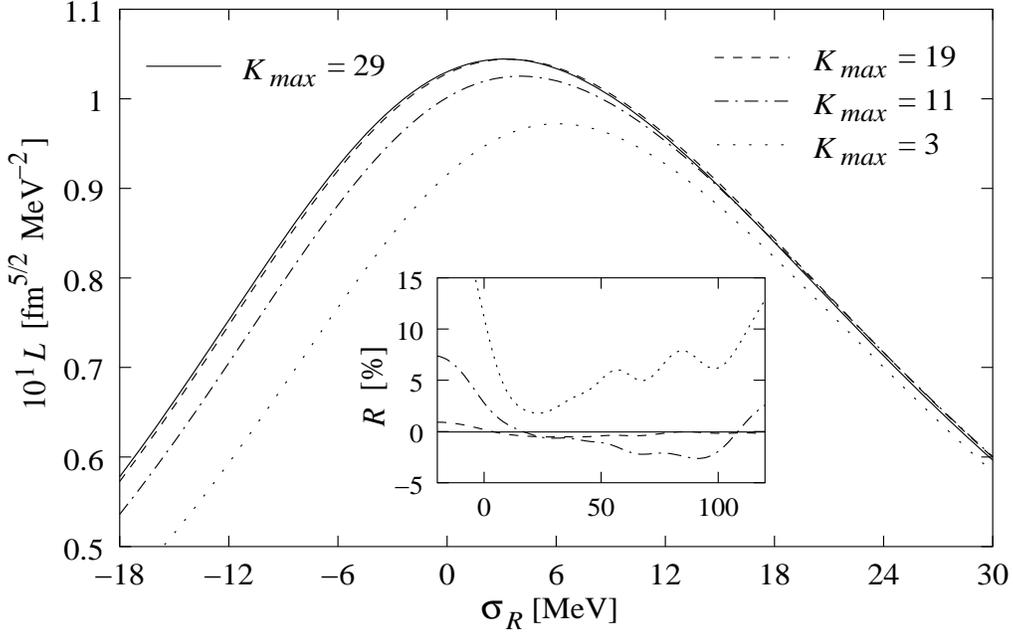}}
\caption{The Lorentz transform of Eq.(\ref{lit}) for $\sigma_I=20$ MeV 
at $E_{n,^3{\rm He}}=$27.9 MeV with various $K_{max}$ values. In the 
inset the deviations $R= 100~[L(K_{max}=29)-L(K_{max})]/L(K_{max}=29)$
are shown.
\label{figure1}}
\end{figure}

In Fig. \ref{figure1}, the convergence of the LIT with respect to 
$K_{max}$ is shown. One sees that for $K_{max}=29$ a good convergence 
is reached.

\begin{figure}
\includegraphics{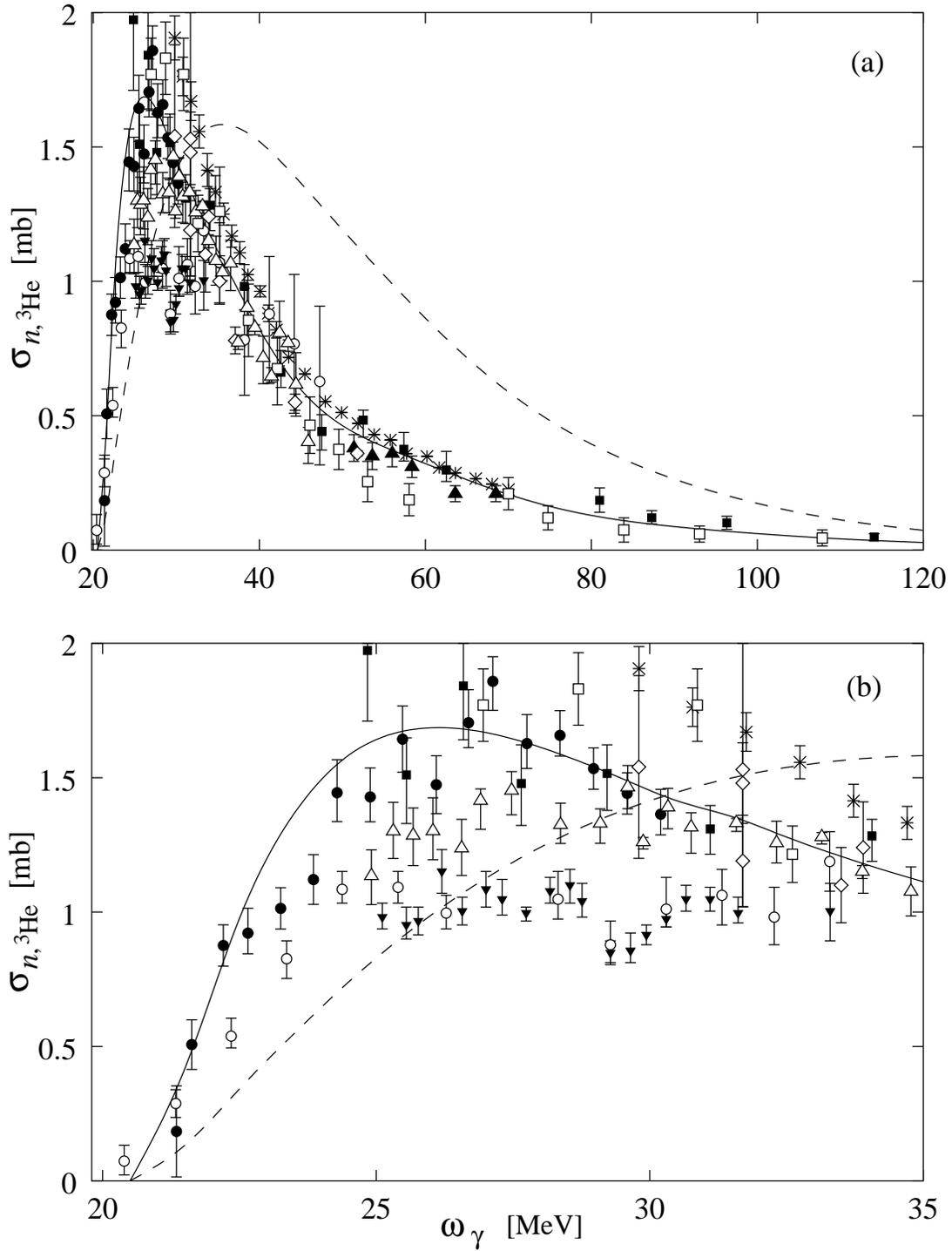}
\caption{$^4$He$(\gamma,n)^3$He cross section up to 120 MeV (a) and 35 MeV (b): full result with FSI included (solid curve), Born 
approximation only (dashed curve); experimental data from \cite{Gor68} (full squares), \cite{Dod72} (diamonds), \cite{Mal73} (stars), \cite{Ark74} (open squares), \cite{Iri75} (full circles), \cite{Bal77} (upward open triangles), \cite{Ber80} (circles),\cite{War81} (downward triangles) and \cite{Sim98} (upward full triangles).
\label{figure2}}
\end{figure}

In Fig. \ref{figure2} we present our results for the 
$^4$He$(\gamma,n)^3$He cross section together with experimental data. 
The difference between the total cross section and its Born approximation 
shows  large effects of FSI. At small energies FSI enhances the 
transition matrix element since the values of the final state continuum 
wave function inside the reaction zone become larger due to the attraction 
between the fragments. At higher energies the transition matrix element 
including FSI becomes smaller than the Born one. This is due to shorter 
wavelengths (higher particle momenta) leading 
to faster oscillations in the integrands. 
 
As it was pointed out in the introduction the experimental results do 
not show a unique picture. In the energy region beyond 35 MeV there is 
an overall agreement of the data of \cite{Gor68,Dod72,Mal73,Ark74,Bal77,%
Ber80,Sim98}, 
whereas in the dipole resonance region big discrepancies are present. 
The data from \cite{Gor68,Mal73,Ark74,Iri75} 
show a rather pronounced dipole peak, while the more recent ones from 
\cite{Ber80,War81} show a flatter behavior (for more 
detailed information see also reviews of ($\gamma,n$) experiments 
\cite{Ber80,CBD83}). 
In the low--energy region our full calculation favours the strongly 
peaked data of \cite{Gor68,Mal73,Ark74,Iri75}.
It is seen from the figure that the higher energy 
experimental cross section agrees quite well with our  full calculation. 
We would like to point out the large effect of FSI in the 
high--energy 
tail. This is probably the region where large dependences of the cross
sections from the details of the force and in particular of the three-body
force are expected (see \cite{ELOT00}) and where it would be 
desirable to have more accurate data.

The present $^4$He$(\gamma,n)^3$He results are at variance with the 
previous 
calculation carried out up to the three--body breakup threshold (AGS 
calculation with a rank 1 separable potential representation of the 
MTI-III potential \cite{Ell96}).

\begin{figure}
\includegraphics{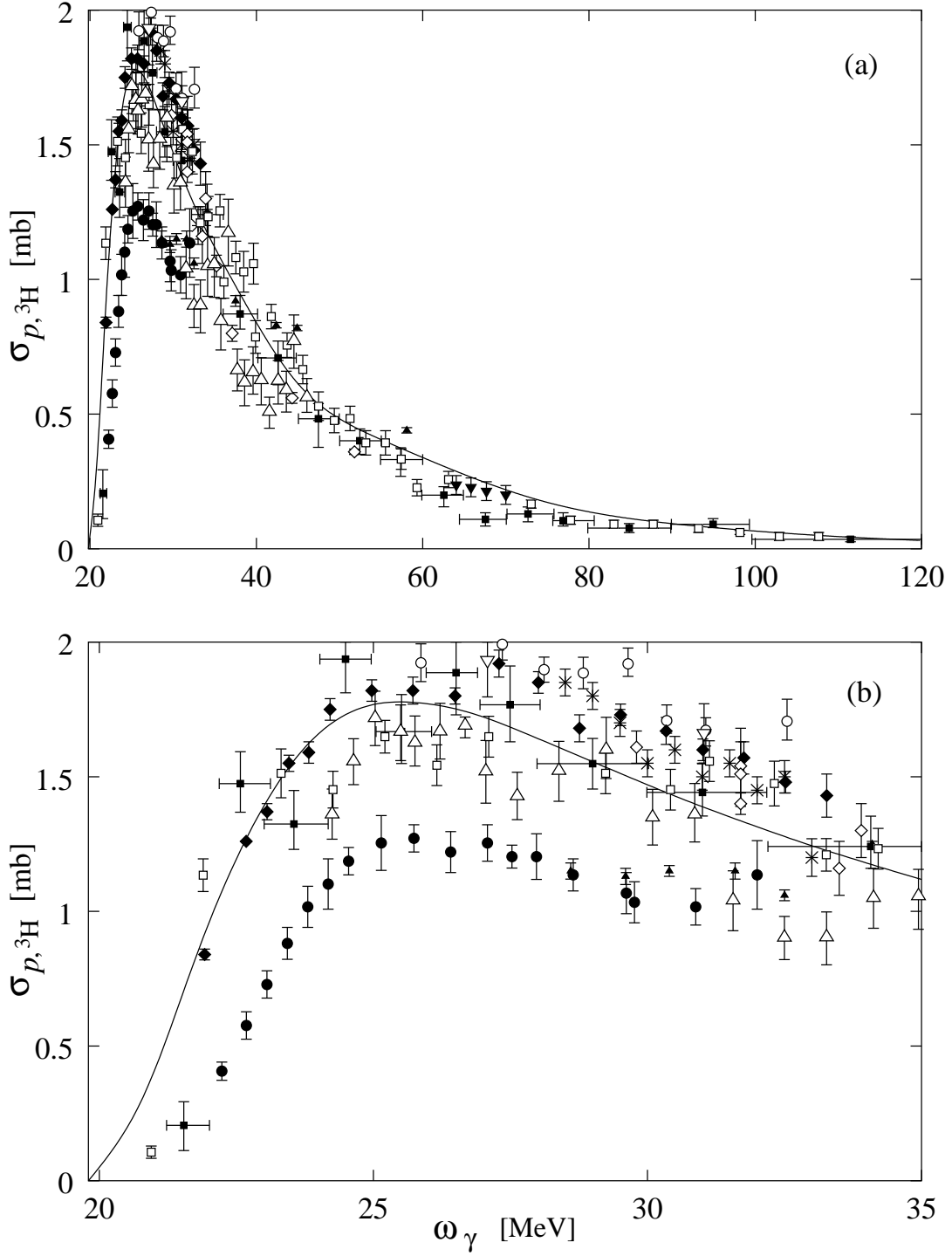}
\caption{$^4$He$(\gamma,p)^3$H cross section up to 120 MeV (a) and 35 MeV (b): full result with FSI included (solid curve); experimental data from \cite{Gor68} (full squares),\cite{Mey70} (full diamonds), \cite{Dod72} (open diamonds), \cite{Ark74} (open squares), \cite{Bal77} (downward open triangles), \cite{McB82} (open circles), \cite{Cal83} (upward open triangles), \cite{Ber88} (upward full triangles), \cite{Fel90} (full circles), \cite{Hoo93} (stars) and \cite{Jon91} (downward full triangles).
\label{figure3}}
\end{figure}

The $^4$He$(\gamma,p)^3$H cross section is compared with data in 
Fig. \ref{figure3}. The experimental situation is similar to the 
$(\gamma,n)$ case considered above. Again some experimental results 
show a pronounced giant dipole peak at low energies, as in case of 
\cite{Gor68,Mey70,Dod72,Ark74,Bal77,McB82,Cal83}, and others present 
much less strength, 
like \cite{Ber88} and \cite{Fel90}. 
At higher energies there is quite a satisfactory agreement among 
experimental data of different groups. As in the $^4$He$(\gamma,n)^3$He 
case, our full results have a large cross section in the dipole resonance
region and favour the strongly peaked data, while beyond the peak one 
finds a rather good agreement between our calculation and all data.
            
\begin{figure}
\includegraphics{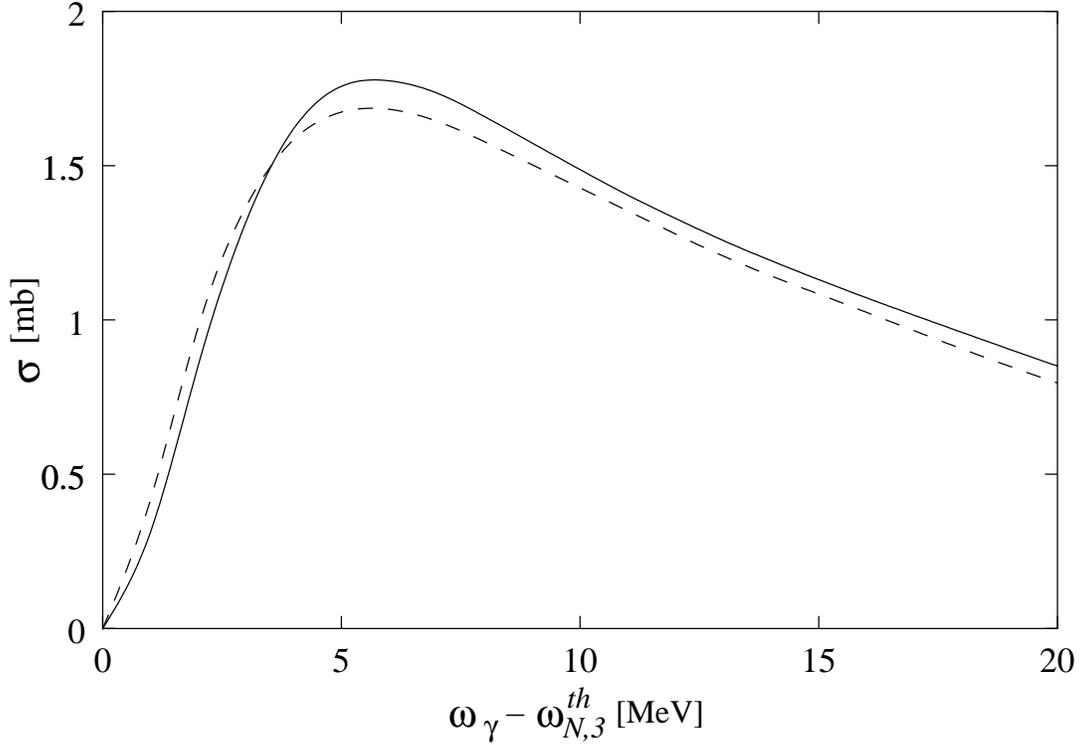}
\caption{$^4$He$(\gamma,p)^3$H (solid curve) and $^4$He$(\gamma,n)^3$He 
(dashed curve) cross sections as a function of photon energy relative 
to the $N,{\mathit 3}$ threshold.
\label{figure4}}
\end{figure}

In Fig. \ref{figure4} we show the effect of the Coulomb interaction in 
the $^4$He$(\gamma,N){\mathit 3}$ reactions. Though the effect is
generally rather small, one notices a suppression
of the cross section near threshold due to the Coulomb interaction 
in the $(\gamma,p)$ case, compensated by an increase at higher energies. 

Since the $^4$He$(\gamma,d)d$ reaction is not induced by the dipole 
operator of (\ref{dipole}), the sum of the $^4$He$(\gamma,p)^3$H and 
$^4$He$(\gamma,n)^3$He cross sections has to be equal to the $^4$He 
total inclusive photoabsorption cross section below the  $n+p+d$ 
breakup threshold. Comparison of the sum of our exclusive cross 
sections in this region with the total inclusive photoabsorption 
cross section calculated independently can therefore serve as a test 
of our results. To this aim, we have calculated also the total 
photoabsorption cross section. This was done using the fact that 
the norm $\left\langle {\widetilde \Psi}_1(\sigma)\left|
\right.{\widetilde\Psi}_1(\sigma)\right\rangle$ represents the LIT of 
this total cross section. Our present CHH calculation 
reproduces the results obtained for the total photoabsorption cross 
section with the effective interaction HH method \cite{BELO01}. 
This can be considered as one more test of our calculation, 
and this is accomplished thanks to the present efforts made in order 
to increase the $K_{max}$ value up to 29.

\begin{figure}
\includegraphics{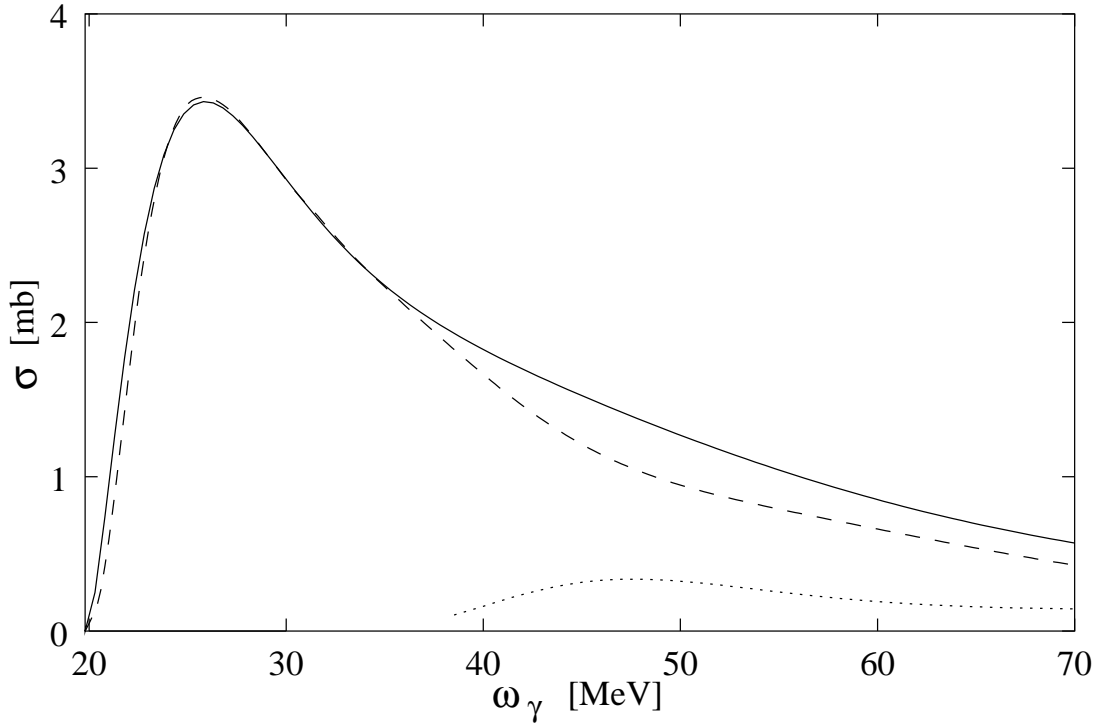}
\caption{Total $^4$He photoabsorption cross section (solid curve) 
compared to the sum of $^4$He$(\gamma,p)^3$H and $^4$He$(\gamma,n)^3$He 
cross sections (dashed curve); more-body break-up cross section obtained 
by subtraction (dotted curve).
\label{figure5}}
\end{figure}

The comparison  of the sum of our cross sections for the 
$^4$He$(\gamma,p)^3$H and $^4$He$(\gamma,n)^3$He reactions with the 
CHH total photoabsorption cross section below the three--body 
breakup threshold is presented in Fig. \ref{figure5}. The agreement 
of the two curves  in this region is quite satisfactory. There is 
only some discrepancy very close to  threshold. The discrepancy could 
presumably be resolved by a calculation of the LIT in the threshold 
region with a smaller $\sigma_I$. However, this also requires an 
increase of the number of hyperradial basis functions along 
with a more precise determination of the nine--dimensional integrals 
(at present Monte Carlo integrations), and thus a considerable improvement 
of the numerical calculation. This uncertainty  is in any
case smaller than the error bars of available experiments in that energy 
range. 

\begin{figure}
\includegraphics{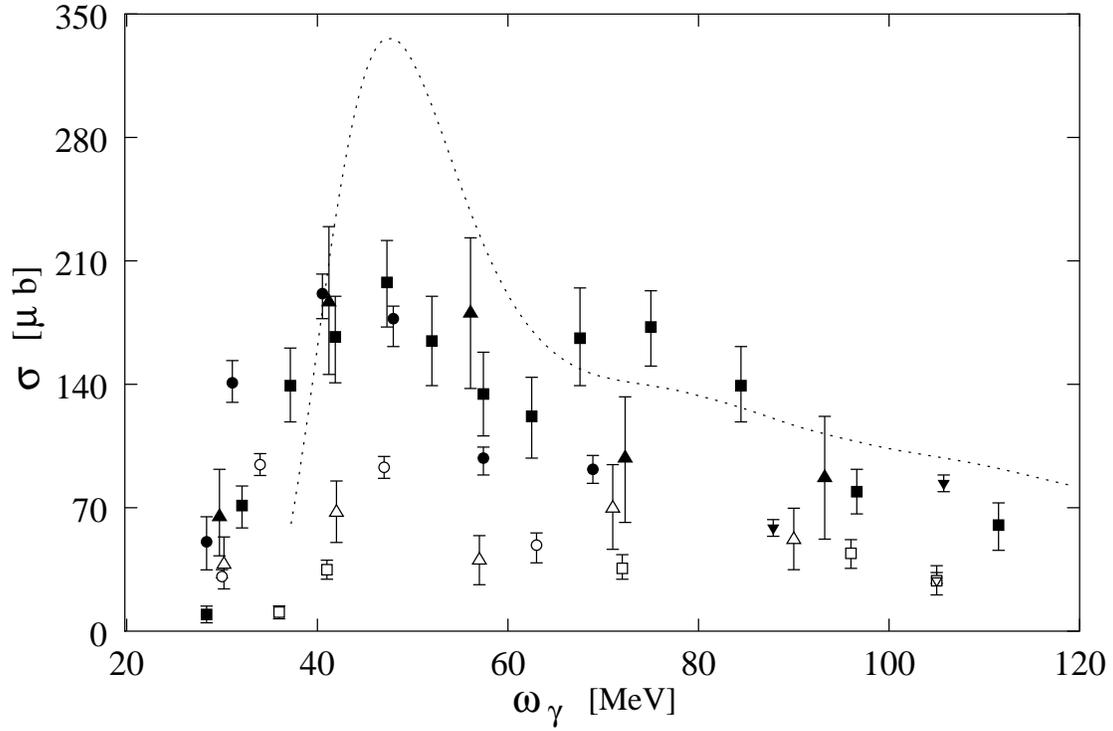}
\caption{More-body break-up cross section (same dotted curve as in Fig. 
\ref{figure5}); $^4$He$(\gamma,pn)d$ (full points) and 
$^4$He$(\gamma,2p2n)$ (open points) experimental data from \cite{Gor58} 
(upward triangles), \cite{Ark70} (squares), \cite{Bal79} (circles), 
\cite{Dor93} (downward triangles).
\label{figure6}}
\end{figure}

Figure \ref{figure5} also shows an indirect determination of 
the 3,4--body breakup cross section, which is obtained by subtraction 
of the $^4$He$(\gamma,p)^3$H and $^4$He$(\gamma,n)^3$He cross sections 
from the total photoabsorption cross section in the region where 
significant differences are found. This more--body breakup cross section 
is shown in Fig. \ref{figure6} together with the three-- and four--body 
disintegration data. Though there are some differences at lower energies 
and the peak of the theoretical cross section is more pronounced, one 
finds an overall fair agreement between theory and experiment. 

In conclusion we summarize our work. We have calculated the total 
photodisintegration cross section of $^4$He in the two mirror channels 
$p,^3$H and $n,^3$He with the MTI-III potential using the LIT method 
and the CHH expansion. For the first time a  microscopic calculation 
for these reactions has been carried out taking fully into account final 
state interaction also beyond the three-- and four--body breakup 
thresholds. The pronounced dipole resonance found in 
the total photonuclear cross sections \cite{ELO97,BELO01}
is almost exhausted by the two-body break up channel. 
The $^4$He$(\gamma,p)^3$H and $^4$He$(\gamma,n)^3$He cross sections  
we obtain have a very similar structure. Both show a pronounced giant 
dipole resonance peak at low energy as obtained in a previous 
calculation of the total $^4$He photoabsorption cross section with the 
same NN potential model. This low--energy behavior of the cross section 
is in agreement with the experimental data of 
\cite{Gor68,Mey70,Mal73,Ark74,Iri75,McB82,Cal83,Hoo93} whereas other 
measurements \cite{Ber80,War81,Ber88,Fel90} show less strength. At 
higher energies there are much less differences among the experimental 
data and one finds a rather good agreement with the theoretical results. 
Our indirect determination of the sum of three-- and four--body breakup 
cross sections shows an overall fair agreement with the sum of three-- 
and four--body break-up data.

We hope to have convinced the reader that further theoretical and 
experimental efforts are necessary for a better understanding of the 
$^4$He photodisintegration. The present theoretical results are 
fully consistent and complete regarding the treatment of the dynamics 
in the initial and final state, though obtained with a simple NN 
potential. In order to confirm the size of the giant dipole peak
one would need a more realistic description of this process 
considering modern realistic NN potentials together with a 
three--nucleon force. 

\subsection*{Acknowledgments}
We would like to thank M.A. Marchisio for supplying us with the 
computer code for the Lanczos algorithm.
S.Q. acknowledges {\bf ECT}$^{\mathbf *}$ for support.
The work of N.B. was supported by the Israel Science Foundation 
(grant no. 202/02). V.D.E. thanks the Department of Physics of the 
University of Trento for support and hospitality.

\end{document}